\pdfoutput=1
\RequirePackage{ifpdf}
\ifpdf % We are running pdfTeX in pdf mode
\documentclass[pdftex]{sigma}
\else
\documentclass{sigma}
\fi

\usepackage{xspace}

\newcommand{\scpr}[2]{\langle#1\, \vert \, #2 \rangle}
\newcommand{\sscpr}[3]{\langle#1\, \vert \, #2 \, \vert \, #3\rangle}

\newcommand{\ket}[1]{\lvert\, #1\rangle}

\newcommand{\comm}[2]{\left[#1\,,\, #2 \right]}

\newcommand{\wt}[1]{\widetilde{#1}}
\newcommand{\hilb}{\mathcal{H}}
\newcommand{\hal}{\hilb_{\text{AL}}}
\newcommand{\bg}[1]{#1^{(0)}}
\newcommand{\pin}{\pi'}
\newcommand{\lqg}{LQG\xspace}

\newcommand{\vac}{\text{vac}}
\newcommand{\emptystate}{\ket{\varnothing}}

\DeclareMathOperator{\abar}{\overline{\mathcal{A}}}

\DeclareMathOperator{\one}{\mathbb{I}}
\DeclareMathOperator{\id}{id}
\DeclareMathOperator{\ad}{Ad}
\DeclareMathOperator{\cyl}{Cyl}
\DeclareMathOperator{\tdiff}{TDif\/f}
\DeclareMathOperator{\diff}{Dif\/f}
\DeclareMathOperator{\symm}{Symm}

\DeclareMathOperator{\hfalg}{\mathfrak{HF}}

\DeclareMathOperator{\gont}{\triangleright}
\DeclareMathOperator{\gone}{\triangleright}
\DeclareMathOperator{\pone}{\triangleright}
\DeclareMathOperator{\pont}{\triangleright}

\begin{document}

\allowdisplaybreaks

\renewcommand{\thefootnote}{$\star$}

\renewcommand{\PaperNumber}{026}

\FirstPageHeading

\ShortArticleName{Loop Quantum Gravity Vacuum with Nondegenerate Geometry}
\ArticleName{Loop Quantum Gravity Vacuum\\ with Nondegenerate Geometry\footnote{This
paper is a contribution to the Special Issue ``Loop Quantum Gravity and Cosmology''. The full collection is available at \href{http://www.emis.de/journals/SIGMA/LQGC.html}{http://www.emis.de/journals/SIGMA/LQGC.html}}}

\Author{Tim KOSLOWSKI~$^\dag$ and Hanno SAHLMANN~$^\ddag$}
\AuthorNameForHeading{T.~Koslowski and H.~Sahlmann}

\Address{$^\dag$~Perimeter Institute for Theoretical Physics, Waterloo, Canada}
\EmailD{\href{mailto:tkoslowski@perimeterinstitute.ca}{tkoslowski@perimeterinstitute.ca}}

\Address{$^\ddag$~APCTP, and Physics Department of POSTECH University, Pohang, Korea}
\EmailD{\href{mailto:sahlmann@apctp.org}{sahlmann@apctp.org}}

\ArticleDates{Received September 23, 2011, in f\/inal form May 03, 2012; Published online May 12, 2012}

\Abstract{In loop quantum gravity, states of the gravitational f\/ield turn out to be excitations over a vacuum state that is sharply peaked on a degenerate spatial geometry. While this vacuum is singled out as fundamental due to its invariance properties, it is also important to consider states that describe non-degenerate geometries. Such states have features of Bose condensate ground states.
We discuss their construction for the Lie algebra as well as the Weyl algebra setting, and point out possible applications in ef\/fective f\/ield theory, Loop Quantum Cosmology, as well as further generalizations.}


\Classification{83C45; 81R15; 46L30; 28C20}

\renewcommand{\thefootnote}{\arabic{footnote}}
\setcounter{footnote}{0}

\section{Introduction}

Observables are quantum mechanically described by operators, and pure states by rays in a~Hilbert space, so the mathematical problem of quantization is in part the representation theory of the underlying operator algebra. The importance of representation theory is in particular due to the fact that dif\/ferent irreducible representations of the same algebra may lead to physically distinct predictions. This can already be observed in the for physicists more familiar case of linear representations of the group ${\rm SU}(2)$. Thus, once a class of operator algebras describing the quantum observables of a system is found, its representation theory is a purely mathematical problem. Progress in this purely mathematical problem can however lead to signif\/icant physical insight.

Strong uniqueness theorems (referred to as F/LOST in the following) state that all regular irreducible representations with dif\/feomorphism invariant vacuum vector of the algebra that underlies Loop Quantum Gravity (LQG) are unitarily equivalent to the familiar Ashtekar--Lewandowski (AL) representation. It turns out that the kinematic vacuum expectation values for all geometric operators, as well as their higher moments, vanish. The vacuum thus represents a degenerate spatial geometry, usually called ``no geometry''. The excitations over this vacuum are one dimensional `f\/lux-tubes' of non-degenerate geometry. It is thus not easy to approximate smooth geometries with them. That said, the AL ground state has the appealing feature that it is the ``no geometry''-state, which is the natural vacuum for background independent quantum gravity, where any geometry is viewed as an excitation of the gravitational f\/ield.

However, the assumptions of invariance under spatial dif\/feomorphisms and irreducibility used in F/LOST are, while natural,
not strictly necessary, at very least not if one is interested in some sort of ef\/fective theory. Thus, dropping these assumptions is interesting and opens the door for new representations, in particular for representations with vacuum vectors that possess nonvanishing expectation values for spatial geometric operators. We review the explicit construction of such representations both for the Lie algebra (i.e.~holonomy-f\/lux algebra) as well as for a slight modif\/ication of the $C^*$-algebra (i.e.~Weyl algebra) that underlies \lqg.

Despite subtle technical dif\/ferences, one f\/inds in both cases the same basic features: There exists a smooth classical background geometry and discrete \lqg excitations on top of the background geometry. This is analogous to states that describe a Bose condensate, where the f\/ield operator attains a nontrivial vacuum expectation value with quantum excitations around this VEV. One may thus view the vacuum vectors of the new representations as condensates of LQG. We conclude the paper by considering applications of the non degenerate representations in ef\/fective f\/ield theory, Wilsonian renormalization and discuss a possible connection with Loop Quantum Cosmology as well as the possibility of yet more general representations.

This review is focused on \cite{Koslowski:2008di,Koslowski:2007kh,Sahlmann:2010hn}. For other non-standard representations with dif\/ferent motivations see~\cite{Dziendzikowski:2009rv, Varadarajan:2007dk}. The technical underpinnings of the AL representation can be found in~\mbox{\cite{Ashtekar:1994wa,Ashtekar:1994mh}}. The F/LOST uniqueness result for the holonomy-f\/lux algebra is based on~\cite{Sahlmann:2002xu,Sahlmann:2002xv,Sahlmann:2003in} and was further developed in \cite{Okolow:2004th, Okolow:2003pk}, and together with the irreducibility result~\cite{Sahlmann:2003qa} resulted in~\cite{Lewandowski:2005jk}. The $C^*$-algebra was developed together with the corresponding irreducibility and uniqueness result in~\cite{Fleischhack:2005km,Fleischhack:2006zs, Fleischhack:2004jc}. For an complementary approach to continuous geometries in LQG from the covariant perspective, see for example~\cite{Oriti:2010hg}.

\section{Algebras underlying Loop Quantum Gravity}

LQG is a canonical quantization program for gravity, based on a formulation in terms of connection variables.
The canonical pair to be quantized consists of an SU(2) connection $A$ and a~densitized triad f\/ield $E$. These f\/ields transform under the automorphisms of
the SU(2) bundle. Upon choosing (local) trivializations, these bundle
automorphisms split into gauge transformations $g$ and dif\/feomorphisms~$\phi$,
\begin{gather*}
 A \overset{g}{\longmapsto} \ad_g(A)+g^{-1}dg, \qquad A
\overset{\phi}{\longmapsto} \phi_{*}A,\\
 E \overset{g}{\longmapsto} \ad_g(E) =:g \gone E, \qquad E
\overset{\phi}{\longmapsto} \phi_{*}E=:\phi \pone E,
\end{gather*}
where $\ad$ is the adjoint action of SU(2) on su(2), and the star signif\/ies
push-forward. The
automorphism group is then the semi-direct product of those subgroups. Poisson brackets are given by
\begin{equation*}
\big\{A^I_a(x),E_J^b(y)\big\}= 8\pi G \beta\delta_J^I \delta(x,y).
%\label{eq:poi}
\end{equation*}
Point f\/ields are usually too singular to give well def\/ined operators in the quantum theory, so a~Poisson algebra of smoother functions on phase space has to be chosen. It is characteristic for \lqg that one considers holonomies
\begin{equation*}
h_{e}[A]=\mathcal{P} \exp \int_{e} A,
\end{equation*}
or more generally, functions of such holonomies as basic functionals of the connection. To be precise, consider functionals of the form
\begin{equation*}
%\label{eq_cyl}
f[A]\equiv F(h_{e_1}[A],h_{e_2}[A],\ldots,h_{e_n}[A])
\end{equation*}
for a f\/inite number of paths $e_1,\ldots, e_n$ forming a graph, and a
function $F$ on $n$ copies of~SU(2). Such
functionals are also called \emph{cylindrical functions}, and the algebra they generate is denoted~$\cyl$.
$\cyl$ carries a natural norm induced by the sup norm for continuous functions on~SU(2). Closure under this norm gives a~C$^*$ algebra
$\overline{\cyl}$ which is isomorphic to $C(\abar)$, the algebra of continuous functions on a set of distributional connections, the so-called \emph{generalized connections}. Dif\/feomorphisms and gauge transformations act as
\begin{gather*}
 (g\gont f)[A]=
F\big(g(e_1(0))^{-1}h_{e_1}g\big(e_1(1), g(e_2(0))^{-1}h_{e_2}g(e_2(1)),\ldots\big)\big),\\
 (\phi\pont f)[A]=
F(h_{\phi(e_1)}[A],h_{\phi(e_2)}[A],\ldots).
\end{gather*}
For the triad f\/ields $E$ on the other hand, the choice of functional is not so clear.
One  natural choice is its \emph{flux} through surfaces $S$:
\begin{equation*}
%\label{eq_flux}
E_{S,r}[E]=\int_S *E_I r^I,
\end{equation*}
where $r$ is a smooth function taking values in su(2)$^*$ and vanishing on $\partial S$, and $*E$ is the two-form
$E^a \epsilon_{abc}\text{d}x^b \wedge \text{d}x^c$.
Another natural choice is to use the imaginary exponentials of those f\/luxes,
\begin{equation*}
w_{S,r}=e^{i E_{S,r}}.
\end{equation*}
These two possibilities lead to slightly dif\/ferent algebras which we will discuss next.

\subsection{Holonomy-f\/lux algebra}

We start with the commutative algebra $\cyl$ and add elements $E_{S,r}$ for surfaces $S$ and maps $r$ to su(2)$^*$ as above. These elements have non-trivial commutation relations with the elements of~$\cyl$ and among each other.
To begin with,
\begin{equation}
\comm{E_{S,r}}{f}=8\pi \beta l^2_P X_{S,r}[f],
\label{eq:ccr}
\end{equation}
which are nothing but the canonical commutation relations. $X_{S,r}[f]$ is an element of $\cyl$ obtained from $f$ by action of a certain derivation $X$.
Then we have
\begin{equation*}
[f,[E_{S_1,r_1}, E_{S_2,r_2}]]=\big(8\pi \beta l^2_P\big)^2[X_{S_1,r_1},
X_{S_2,r_2}][f]
\end{equation*}
and similar higher order relations. These are consequences of~\eqref{eq:ccr} and the requirement that the commutator in the resulting algebra satisfy the Jacobi identity.

There are also relations, concerning linearity in the $r_i$ and related properties, which we will not discuss in detail here. Finally, the algebra becomes a $*$-algebra by setting $E_{S,r}^*=E_{S,r}$ for real $r$. We will denote this algebra with~$\hfalg$. Dif\/feomorphisms and gauge transformations act as
automorphisms of $\hfalg$,
\begin{equation*}
(g\gone E_{S,r})=E_{S,r\circ \ad_{g^{-1}}},\qquad
(\phi\pone E_{S,r})=E_{\phi(S),\phi_*r}.
\end{equation*}
A more precise def\/inition of $\hfalg$ can be found for example in~\cite{Lewandowski:2005jk}.

\subsection[$C^*$-algebra (Weyl algebra)]{$\boldsymbol{C^*}$-algebra (Weyl algebra)}

Given a $C^*$-algebra\footnote{A $C^*$-algebra over $\mathbb C$ is an algebra $\mathfrak A$ that is also Banach space w.r.t.\ a norm $||\cdot||:\mathfrak A\to \mathbb R^+_o$ and an involution $*:\mathfrak A\to\mathfrak A$, s.t.\ $||a^*a||=||a||^2$ for all $a\in \mathfrak A$.} $\mathfrak A$ and a state\footnote{A state is a continuous linear functional satisfying positivity $\omega(a^*a)\ge 0$ for all $a\in \mathfrak A$, as well as satisfying $\omega(1)=1$ if $\mathfrak A$ is unital.} $\omega: \mathfrak A \to \mathbb C$, one can construct a Hilbert space and a~continuous $*$-representation of the $C^*$-algebra through the GNS construction, which involves the following steps:
\begin{enumerate}\itemsep=0pt
 \item Consider the quotient of $\mathcal H^o_\omega=\mathfrak A / \mathcal I_\omega$, where $\mathcal I_\omega:\{a\in \mathfrak A: \omega(a^*a)=0\}$ denotes the Gel'fand ideal.
 \item Quotienting by $\mathcal I_\omega$ def\/ines the map $\eta_\omega: \mathfrak A \to \mathcal H^o_\omega$, which def\/ines a representation $\pi_\omega$ of $\mathfrak A$ on $\mathcal H_\omega^o$ by $\pi_\omega(a): \eta_\omega(b) \mapsto \eta_\omega(ab)$ for all $a,b\in \mathfrak A$.
 \item Def\/ine the inner product $\langle\cdot,\cdot\rangle_\omega:\mathcal H_\omega^o\times\mathcal H_\omega^o \to \mathbb C$ by $\langle \eta_\omega(a),\eta_\omega(b)\rangle_\omega:=\omega(a^*b)$ and complete~$\mathcal H_\omega^o$ to a~Hilbert space $\mathcal H_\omega$ in this inner product.
 \item Extend the representation $\pi_\omega$ by continuity from $\mathcal H_\omega^o$ to $\mathcal H_\omega$.
\end{enumerate}
If $\mathfrak A$ has a unit then $\eta_\omega(1)$ is a cyclic vector $\Omega_\omega$; moreover any representation $(\pi,\mathcal H)$ of $\mathfrak A$ with cyclic\footnote{$v$ is cyclic if $\{\pi(a)v\}_{a \in \mathfrak A}$ is dense in $\mathcal H$.} normalized vector $v\in \mathcal H$ def\/ines a state $\omega_v$ def\/ines by $\omega_v(a):=\langle v,\pi(a) v\rangle_{\mathcal H}$ and the GNS-representation constructed from $\omega_v$ is unitarily equivalent to $(\pi,\mathcal H)$, so there is a one to one correspondence between cyclic representations and states.

To apply this framework to \lqg, Fleischhack considers a class of $C^*$-algebras that naturally includes the Weyl algebra of \lqg. The class of algebras he considers is constructed from structure data given by a base manifold $\mathbb X$, a compact gauge group $\mathbb G$, a groupoid $\mathcal P$ of paths in $\mathbb X$ of a~certain smoothness class, a subset $\mathcal S$ of the quasi-surfaces in $\mathbb X$ and for each $S\in \mathcal S$ a subset $\Sigma(S)$ of their intersection functions and a subset $\Delta(S)$ of functions $\mathbb X\to \mathbb G$, as well as a subset $\mathcal D$ of the graphomorphisms that leave $\mathcal S$ invariant and act covariantly on~$\Sigma$ and~$\Delta$. Fleischhack then uses additional restrictions (as optimal data) on the structure data that ensure that the constructions of the F/LOST theorem hold. This abstract structure data can be used to construct a generalization of the algebra that underlies \lqg; we will now illustrate this construction.
\begin{enumerate}\itemsep=0pt
 \item The conf\/iguration space of \lqg consists of generalized connections, which are morphisms from the path groupoid $\mathcal P$ to the gauge group $\mathbb G$. This is a generalization of the morphisms def\/ined by gauge f\/ield conf\/igurations given by $A: e\in \mathcal P\mapsto h_e(A)\in \mathbb G$, where $h_e(A)$ denotes the parallel transport along $e$. Using the partial order among graphs $\gamma$ (as f\/inite subgroupoids $\mathcal P_\gamma$ of the path groupoid $\mathcal P$) given by the subgraph relation, one can equip the space $\mathcal A$ of the morphisms $A$ with the Tychonov topology inherited from the compactness of $\mathbb G$. The algebra $C(\mathcal A)$ of continuous functions in this topology is the $C^*$-completion of continuous functions on morphisms $\mathcal A_\gamma \ni A_\gamma:\mathcal P_\gamma \to \mathbb G$. Functions that are determined by the dependence on the image of a f\/inite number of elements of $\mathcal P$ are referred to as cylindrical functions; these are dense in $C(\mathcal A)$.
 \item Dif\/feomorphisms $\phi \in \mathcal D$ act by their pull-backs, so $\phi$ maps $A:e \mapsto h_e(A)$ to $\phi\triangleright A: e\mapsto h_{\phi(e)}(A)$. Gauge transformations $\Lambda: \mathbb X \to \mathbb G$ act on $A: e\mapsto h_e(A)$ by $\Lambda\triangleright A: e \mapsto \Lambda^{-1}(e(0)) h_e(A) \Lambda (e(1))$, where $e(0)$, $e(1)$ denote the initial resp.\ f\/inal point of $e$. Both sets of transformations act as natural transformations of the functors $A$.
 \item The (uniform) Ashtekar--Isham--Lewandowski measure $d\mu_{AIL}$ on $\mathcal A$ is def\/ined as the measure whose push-forward under the projections from $\mathcal A$ onto $\mathcal A_\gamma$ coincides with the product of Haar measures of $\mathbb G$ on each edge $e \in \gamma$. This allows us to construct a Hilbert space $\mathcal H_{AL}=L^2(\mathcal A,d\mu_{AIL})$ on which the algebra of cylindrical functions (and its $C^*$-completion) act as multiplication operators. Moreover, since $d\mu_{AL}$ is invariant under the pull-back action of dif\/feomorphisms and gauge-transformations, one can def\/ine unitary operators on~$\mathcal H_{AL}$ acting as $U_\phi: f(A) \mapsto f(\phi \triangleright A)$ and $U_\Lambda: f(A) \mapsto f(\Lambda \triangleright A)$ for $\phi \in \mathcal D$ and $\Lambda \in \Delta$.
 \item The def\/inition of the exponentiated momentum (f\/lux) operators is subtle: An oriented surface $S$ def\/ines an intersection function $\sigma(S,e)$ vanishing for edges $e$ completely within or completely outside $S$ and $\pm 1$ for edges beginning resp.\ ending above resp.\ below $S$. Moreover, if both $S$ and $e$ are restricted to the piecewise analytic category then every edge $e=e_1\circ \cdots \circ e_n$ can be f\/initely decomposed into $S$-admissible pieces $e_i$ that are either completely inside, outside or beginning or ending on $S$. Let $\Lambda \in \Delta(S)$ then we def\/ine the translation $\Theta_{\Lambda,S,\sigma}$ on $\mathcal A$ by $\Theta_{\Lambda,S,\sigma} A: e_i \mapsto \Lambda^{-\sigma(S,e_i)} (e_i(0))h_e(A)\Lambda^{\sigma_{S,e_i}}(e_i(1))$, where the pieces $e_i$ are assumed to be $S$-admissible. We can now def\/ine the Weyl operators corresponding to exponentiated f\/lux operators as the pull-backs under these translations: $w_{\Lambda,S,\sigma}:=\Theta^*_{\Lambda,S,\sigma}$. It turns out that these Weyl operators leave the Ashtekar--Isham--Lewandowski measure invariant and can thus be extended to unitary operators on $\mathcal H_{AL}$. There is however and additional subtlety: Products of Weyl operators through natural surfaces require us to include lower dimensional ``quasisurfaces'' $S$ and non-natural intersections functions $\sigma$ in the group $\mathcal W$ generated by the natural Weyl operators $w_{\Lambda,S,\sigma}$.
 \item With these preparations we def\/ine the $*$-algebra $\mathfrak A_o$ as the algebra of f\/inite sums of ordered products $a=\sum_{i=1}^n f_i\circ w_i$, where $f_i\in C(\mathcal A)$ and $w_i\in\mathcal W$. Using the the action of~$C(\mathcal A)$ as multiplication operators and $\mathcal W$ as pull-backs under translations on $\mathcal H_{AL}$ we have a~canonical representation of $\mathfrak A_o$ as bounded operators on $\mathcal H_{AL}$. This allows us to def\/ine the completion $\mathfrak A$ as the smallest $C^*$-algebra in $\mathcal B(\mathcal H_{AL})$ that contains $\mathfrak A_o$.
\end{enumerate}
The operators $f\in C(\mathcal A)\subset \mathfrak A$ are the conf\/iguration operators and the operators \mbox{$w\in \mathcal W\subset \mathfrak A$} are the exponentiated momentum operators of \lqg and the dif\/feomorphisms and gauge transformations have same action as in \lqg. To obtain the usual (unbounded) f\/lux operators, one has to take the derivative at the unit element of $\mathcal W$, which imposes a continuity condition on the representation of $\mathfrak A$. This justif\/ies calling~$\mathfrak A_o$ the Weyl algebra underlying \lqg.

\subsection{F/LOST representation theorem}

The AL-representations that we have described above are \emph{fundamental}  for LQG in a specif\/ic sense: They are the unique representations with certain natural properties. This is the content of the uniqueness results \cite{Fleischhack:2004jc,Lewandowski:2005jk, Okolow:2003pk,Sahlmann:2002xv,Sahlmann:2003in}. Roughly speaking, the properties that specify the representations uniquely are the following:
\begin{enumerate}\itemsep=0pt
	\item \emph{Cyclicity.} There exists a vector $\Omega$ in the representation Hilbert space such that application of all the operators of the algebra creates a set of vectors dense in the Hilbert space.
	\item \emph{Invariance.} The cyclic vector $\Omega$ is invariant under dif\/feomorphisms.
\end{enumerate}
We should point out that in the case of the holonomy-f\/lux algebra, the requirement of cyclicity presupposes that $\Omega$ is in the domain of all the operators representing $\hfalg$, and it implies that there is a common dense domain for all of them.
In the case of the Weyl-like algebra $\mathfrak{A}$ one requires additionally
\begin{enumerate}\itemsep=0pt
\setcounter{enumi}{2}
  \item \emph{Regularity.} The representation of the Weyl elements $w\in\mathcal{W}$ has to be continuous in the smearing function.
\end{enumerate}
Cyclicity is natural in that it follows from irreducibility. The cyclic representations are thus the simple building blocks out of which one can build more complicated representations. Another way to motivate cyclicity is to view it as a replacement of the lowest energy property of ground states in the absence of a good notion of energy.

Invariance is a natural requirement given the fact that the constraints of canonical gravity in the Ashtekar--Barbero variables require, among other things, invariance under gauge transformations and spatial dif\/feomorphisms. In fact, the AL vacuum turns out to be a \emph{solution} of all the constraints of LQG. Moreover cyclicity and invariance together imply that gauge transformations and spatial dif\/feomorphisms are implemented by unitary operators. This in turn greatly facilitates f\/inding general solutions to the constraints via group averaging (see for example~\cite{Ashtekar:1995zh}).

Regularity insures that the f\/luxes themselves are available in the representation, thus ma\-king it possible to obtain geometric operators and facilitate quantization of the Hamiltonian constraint.

It should be noted that while the conditions above are the main ones that lead to uniqueness, there are further technical conditions in the uniqueness theorems. For details we refer to the original literature. Perhaps more importantly, we want to emphasize that while the requirements are natural, they are not absolutely necessary. This is due to the fact that the algebras and representations under consideration are kinematic by nature. The algebra elements are generically not observables, and the states generically do not solve the constraints. Rather, these representations are a departure point for constructing physical states and observables. If other ways to achieve this can be found, they may be equally viable, or at least interesting in certain contexts. It is in this spirit that we will drop the invariance condition and construct representations with ground states that describe non-degenerate geometries in the next section.

%%%%%%%%%%%%%%%%%%%%%%%%%%%%%%%%%%%%%%%%%%%%%%%%%%%%%%%%%%%%%%%%%%%%
\section{Nondegenerate vacua and representations}
%%%%%%%%%%%%%%%%%%%%%%%%%%%%%%%%%%%%%%%%%%%%%%%%%%%%%%%%%%%%%%%%%%%%

The cyclic state of the AL-representation is an eigenstate of all the f\/lux operators (and, more generally, of all geometric operators) with eigenvalue zero. Conversely, it maximizes quantum mechanical uncertainty in the holonomies in a precise technical sense. Thus this state treats the canonical variables quite asymmetrically. On a fundamental level, some justif\/ication for this is provided by the uniqueness theorems that were brief\/ly described above. But in order to more ef\/fectively describe non-degenerate geometries in the quantum theory,  there have long been ef\/forts to f\/ind states that divide uncertainty more evenly, and give the f\/lux operators non-zero expectation values. One strategy is to seek states with such properties in $\hal$, leading to weave states \cite{Ashtekar:1992tm} and coherent states \cite{Bahr:2007xa,Bahr:2007xn,Bianchi:2009ky,Freidel:2010tt,Thiemann:2002vj, Thiemann:2000bw}. These states achieve the stated goals for a f\/inite number of degrees of freedom, but there are dif\/f\/iculties to extend this to inf\/initely many degrees of freedom. A quick way to see the source of these dif\/f\/iculties is the following consideration\footnote{It was developed together with M.~Varadarajan.}.
Instead of trying to construct a smoothly peaked function, let us f\/irst consider characteristic functions
\begin{equation*}
\chi_M(A):=\begin{cases}1,& A \in M,\\
0, & \text{otherwise} \end{cases}
\end{equation*}
for some subset $M\subset\abar$. Pick a neighborhood $M_{g_0}$ around a point $g_0$ in SU(2) and let $M_{e,g_0}=\{A\in\abar: h_e(A)\in M_{g_0}\}$ for some edge $e$. Then it is easy to show that
\begin{equation*}
\|\chi_{M_{e,g_0}}\|=\text{Vol}(M_{g_0}),
\end{equation*}
where Vol denotes the volume on the group. This shows that the norm of this characteristic function can get small, despite the fact that the set $M_{e,g_0}$ is huge. Restricting further holonomies to the neighborhood $M_{g_0}$ results in
\begin{equation*}
\|\chi_{M_{e_1,e_2,\ldots, e_n,g_0}}\|=\text{Vol}(M_{g_0})^n.
\end{equation*}
From this it is clear that to obtain something that is peaked on inf\/initely many edges, one has to rescale the characteristic functions and take a limit, and that this limit \emph{can not be a function} anymore.

Indeed there exist objects that have the properties of Gaussians outside of $\mathcal{H}_{\text{AL}}$, namely Gaussian \emph{measures} on $\abar$. Measures on $\abar$ are generally given by families of measures on cylindrical subspaces, fulf\/illing certain consistency conditions \cite{Ashtekar:1994wa,Ashtekar:1994mh}. For Gaussian measures \cite{Ashtekar:2001xp,Sahlmann:2002xv,Thiemann:2002vj,Varadarajan:1999it}, the family has the form
\begin{equation*}
\text{d}\mu_\gamma = e^{-G_\gamma{}^{IJ}_{ee'}X_I^eX^{e'}_J}\delta^\gamma_{z} \, \text{d}\mu_\text{Haar},
\end{equation*}
where $G_\gamma$ is a \emph{graph metric}, the $X$ represent invariant vector f\/ields acting on the variables corresponding to the various edges of the graph $\gamma$, and the delta-function is
$\delta^\gamma$ is peaked at a~point $z$ in the complexif\/ication of SU(2)$^{|\gamma|}$.  Consistency of the family translates into conditions on the family of $G_\gamma$, and some solutions are known~\cite{Ashtekar:2001xp}. The problem with these measures is that, while they furnish new representations for $\cyl$, they generally do not support representations of the f\/luxes~\cite{Sahlmann:2002xv}. The reason is that to obtain symmetric operators, the f\/luxes have to acquire a~measure dependent divergence term,
\begin{equation*}
\pi(E_{S,f})=X_{S,f}+ \frac{i}{2}\text{div}_\mu(S,f).
%\label{eq:div}
\end{equation*}
$\text{div}_\mu(S,f)$ is formally an operator which preserves the commutation relations and compensates for the failure of $X$ to be symmetric, i.e., $i\text{div}_\mu(S,f)=X^\dagger_{S,f}-X_{S,f}$.
For the known measures, the divergence term is so singular that a representation of the f\/luxes by symmetric operators simply does not exist~\cite{Sahlmann:2002xv}. Thus we do not know any states that divide uncertainty more evenly between more than a f\/inite number of canonical pairs. But while we do not know how to broaden the delta-function-like peak of the AL ground state
it turns out that it \emph{is} possible to \emph{shift} this peak to non-degenerate geometries. This is already a big step in and of itself, because the resulting representations describe excitations over a geometric background. But it may only be the f\/irst step towards some form of coherent states that more accurately represent the observed geometric condensate that we classically describe as a space-time geometry. The states that we review more precisely below, contain a condensate of \emph{spatial} geometry, but the extrinsic curvature of the spatial slice, and hence the \emph{space-time} geometry is highly f\/luctuating and quantum mechanical in these states.
We will now turn to a more detailed description.

\subsection{Representations of the holonomy-f\/lux algebra\\ with nondegenerate background geometry}

Now we will discuss the representations of $\hfalg$ with nondegenerate background geometry~\cite{Koslowski:2007kh, Sahlmann:2010hn}. Let $\bg{E}$ be
a classical triad f\/ield. Then we can def\/ine a \emph{new}
representation $\pin$ on the AL Hilbert space $\hal$, by changing the action of the f\/luxes:
\begin{equation}
 \pin(E_{S,r})={X}_{S,r} +  \bg{E}_{S,r}\id, \qquad \text{with}\quad
\bg{E}_{S,r}=\int_S *\bg{E}{}^I r_I.
\label{eq:nflux}
\end{equation}
It is easily checked that this gives another representation of the algebra $\hfalg$. We
list some elementary properties.
\begin{enumerate}\itemsep=0pt
 \item The spectra of the f\/luxes have changed. If $\lambda$ is an eigenvalue
of $X_{S,f}$ then $\lambda + \bg{E}_{S,f}$ is an eigenvalue of $\pin(E_{s,f})$.
In particular, the AL vacuum $\emptystate$ is now
an eigenstate of $\pin(E_{s,f})$ with eigenvalue $\bg{E}_{S,f}$, which is
non-vanishing in general.
\item The new representation is still cyclic, with $\emptystate$ as cyclic
vector.
\item The new representation is unitarily inequivalent to the AL representation.

\item The standard kinematic representation can be viewed as a special case
of the new one, for $\bg{E}=0$.
\end{enumerate}
1~is immediate since the new term is proportional to the identity and hence all states are its eigenstates. 2~can be shown by making use of the fact that the unit operator is part of the algebra. It can be used to remove the new terms in~\eqref{eq:nflux} and show that old and new representation generate the same dense subspace upon acting on $\emptystate$. As for 3, the new representation is unitarily inequivalent, because a unitary transformation can not change the spectra. 4~is obvious.

To better interpret the new representations, it is very useful to consider the action of operators corresponding to geometric measures such as area and volume. In the AL-representation, these are def\/ined by giving a regularized expression in terms of the elementary variables, quantizing this regularized expression, and then taking the cutof\/f to zero in the quantum theory~\cite{Ashtekar:1996eg,Lewandowski:1996gk}. A~priori, it is not clear if this strategy still works in the new representation since, due to the extra term in~\eqref{eq:nflux}, the regularized operators contain additional terms, the number of which tends to inf\/inity as the regulator is removed. It turns out, however, that all goes well.
\begin{proposition}[\cite{Sahlmann:2010hn}]
The standard regularization and quantization procedures for area and
volume operators can be applied to the new representations and lead to well
defined operators. We find
\begin{equation*}
 V_{R}=V_R^{\rm vac}+ \bg{V}_R\id, \qquad A_S=A^{\rm vac}_S+\bg{A}_S\id
\end{equation*}
with $V^{\rm vac}$, $A^{\rm vac}$ the geometric operators in the vacuum representation,
and
$\bg{V}_R$, $\bg{A}_S$ the classical values in the background geometry.
\end{proposition}
This shows clearly that the new representations contain a background geometry given by the background f\/ield~$\bg{E}$.

We turn next to the gauge symmetries of LQG: The operators implementing dif\/feomorphisms and gauge transformations in
the standard representation are still well def\/ined and unitary in the new
representation, but it can be easily checked that they do not implement the
algebra-automorphisms anymore. For example
\begin{gather*}
 U_\phi \pin(E_{S,f}) U_{\phi^{-1}}=
U_\phi \big(X_{S,f}+ \bg{E}_{S,f}\one \big) U_{\phi^{-1}}
= X_{\phi(S),\phi_*f} + \bg{E}_{S,f}\one
\neq \pin(\phi \pone E_{S,f}),
\end{gather*}
by virtue of the fact that the $U_\phi$ act on the $X_{S,f}$ in the standard
way, but commute with the new c-number term. In other words, the problem is that the~$U_\phi$ do not change the background geometry.

This can be remedied by going over to a large direct sum of Hilbert spaces. For a given background~$\bg{E}$ we consider
\begin{equation}
\label{eq_dirsum}
\mathcal{H}_{G(\bg{E})}:=\bigoplus_{\bg{\wt{E}}\in G(\bg{E})}
\mathcal{H}_{\bg{\wt{E}}},
\end{equation}
where $G(\bg{E})$ is the equivalence class of $\bg{E}$ with respect to
dif\/feomorphisms and gauge transformations. The Hilbert space is thus
ef\/fectively labeled by a spatial background metric modulo dif\/feomorphisms.
We will write $\ket{T,\bg{\wt{E}}}$
for the cylindrical state $T$ in the (new) representation with background
geometry $\bg{\wt{E}}$ in the above Hilbert space. Thus we have
\begin{equation*}
%\label{eq_scpr}
\scpr{T,\bg{E}}{T',\bg{E'}}=\scpr{T}{T'}\delta_{\bg{E},\bg{E'}},
\end{equation*}
where the f\/irst scalar product on the right hand side is just the one in the
vacuum representation.
The Hilbert space $\mathcal{H}_{G(\bg{E})}$ carries a representation of $\hfalg$, which is simply the direct sum of the representations on the
individual spaces. But now one can implement the dif\/feomorphisms and gauge
transformations unitarily, by setting \cite{Koslowski:2007kh,Sahlmann:2010hn}
\begin{equation*}
U_\phi\ket{T,\bg{E}}=\ket{\phi \pont T, \phi \pone \bg{E}}, \qquad
U_g\ket{T,\bg{E}}=\ket{g\gont T, g \gone \bg{E}}.
\end{equation*}
It is easy to show that with this def\/inition, the transformations are also
correctly implemented on the operators. A straightforward calculation shows that not only the groups of gauge
transformations and spatial dif\/feomorphisms are correctly implemented, but also
their direct product.

The unitary implementation of the symmetries can now be used to f\/ind invariant states and Hilbert spaces by \emph{group averaging}, just as has been done for the vacuum representation \cite{Ashtekar:1996eg}. Let us consider the case of dif\/feomorphisms as example.We will call
a dif\/feomorphism $\phi$ a symmetry of a triad $\bg{E}$, if $\phi$
leaves $E$ invariant, $\phi \pone \bg{E}=\bg{E}$. Then
\begin{enumerate}\itemsep=0pt
\item Let $\diff$ be the group of all dif\/feomorphisms.
\item Let $\diff_{(\alpha,\bg{E})}$ be the group of dif\/feomorphisms that are
symmetries of $\bg{E}$ and map the graph $\alpha$ onto itself.
\item Let $\tdiff_{(\alpha,\bg{E})}$ be the group of dif\/feomorphisms that are
symmetries of $\bg{E}$ and map each edge  of the graph $\alpha$ onto itself.
Note: Again, if $\bg{E}=0$ these are just the dif\/feomorphisms mapping each edge
of $\alpha$ onto itself, denoted as $\tdiff_\alpha$ in \cite{Ashtekar:1996eg}.
\item Let $\symm_{(\alpha,\bg{E})}$ be the quotient
$\diff_{(\alpha,\bg{E})}/\tdiff_{(\alpha,\bg{E})}$.
\end{enumerate}
Note that these def\/initions closely follow \cite{Ashtekar:1996eg}. Let now $T_\alpha$ be a cylindrical function that depends on the holonomies of all
the edges of $\alpha$ in a non-trivial way. We can then def\/ine the linear form
$(T_\alpha,\bg{E}|$ on $\mathcal{H}_{G(\bg{E})}$ as follows:
\begin{equation*}
%\label{eq_diff}
(T_\alpha,\bg{E}|T_\beta, \bg{E'}\rangle
:=\sum_{G(\phi)\in\diff/\diff_{(\alpha,\bg{E})}}
\sscpr{T_\alpha,\bg{E}}
{P_{(\alpha,\bg{E})} U^\dagger_\phi}
{T_\beta, \bg{E'}},
\end{equation*}
where the projection $P_{(\alpha,\bg{E})}$ is def\/ined as
\begin{equation*}
%\label{eq_pro}
P_{(\alpha,\bg{E})}
\ket{T_\alpha,\bg{E}}:=\frac{1}{\big|\symm_{(\alpha,\bg{E})}\big|} \sum_{G(\phi)\in
\symm_{(\alpha,\bg{E})}} U_\phi\ket{T_\alpha,\bg{E}}.
\end{equation*}
Now it is easy to show that
\begin{proposition}
The linear functionals $(T_\alpha,\bg{E}|$ are well defined, finite, and diffeomorphism invariant,
\begin{equation*}
(T_\alpha,\bg{E}|\circ U_\phi=(T_\alpha,\bg{E}|\qquad \text{for all} \quad \phi\in\diff.
\end{equation*}
\end{proposition}
Similar results can be obtained for gauge transformations and, taking the semidirect product of dif\/feomorphisms and gauge transformations, for bundle automorphisms. One thereby obtains the automorphism invariant Hilbert space $\mathcal{H}_\text{aut}$. Thus the quantum kinematics can be developed to exactly the same point for the new representations as for the AL representation. What is more, the latter appears simply as a special case of the constructions we sketched above.

We should stress that, while $H_{G(\bg{E})}$ is very large, this is partially remedied by the group averaging. Vectors
$\ket{1,\bg{E}}$, $\ket{1,\bg{E'}}$ $\in H_{G(\bg{E})}$ are mapped onto \emph{the same} vector in $\mathcal{H}_\text{aut}$. More generally
\begin{equation*}
(f,\phi \pone E|=(\phi^{-1}\pont f,E|.
%\label{eq:}
\end{equation*}
We should also say that it seems that there is no problem with operators on $\mathcal{H}_\text{aut}$. As an example consider the operator $V$ for the volume of the entire spatial slice. Also in the new representations, it commutes with all automorphisms. It thus def\/ines an operator on $\mathcal{H}_\text{aut}$. Moreover, this operator acts in precisely the way one would expect. If $f$ is an eigenstate of $V^{\vac}$, with eigenvalue $\lambda$, then  $|f,\bg{E})$ is an eigenvector of $V$ with eigenvalue $\lambda+\bg{V}$.

\subsection{Nondegenerate representations of the Weyl algebra}

Let us comment on the structure of the Weyl algebra of \lqg, before we construct nondegenerate representations of a slight modif\/ication of $\mathfrak A_o$. We have a compact conf\/iguration space~$\mathcal A$ and a~group $\mathcal W$ of transformations acting on this space. Given such data one can build the a $*$-algebra of f\/inite sums of the form $a=\sum_{i=1}^n f_i \circ w_i$, where $f_i \in C(\mathcal A)$ and $w_i \in \mathcal W$. The multiplication is given by linearity and
$(f_1\circ w_1) (f_2 \circ w_2) = f_1 \alpha_{w_1}(f_2)\circ w_1 w_2$ and the involution is given by antilinearity and $(f\circ w)^*=\alpha_{w^{-1}} (\overline f) \circ w^{-1}$, where $\alpha_w$ denotes the action of the Weyl operators as pull-backs. Given a normalized $\mathcal W$-invariant measure $d\mu$ on $\mathcal A$, one f\/inds by direct computation that
\begin{equation}\label{equ:AL-state}
 \omega_o(f\circ w):=\int d\mu(A) f(A)
\end{equation}
is a bounded positive linear functional on $\mathfrak A_o$ and can thus be used for the GNS construction, which yields the standard Hilbert space representation of \lqg. The def\/inition~(\ref{equ:AL-state}) can be ge\-ne\-ralized in various ways; let us e.g.\ assume we have a character\footnote{A character of a group $G$ is a continuous morphism from $G$ to $U(1)$.} $\chi$ of the group $\mathcal W$, then we can consider the functional obtained by the linear extension of
\begin{equation}\label{equ:character-state}
     \omega_\chi(f\circ w):=\chi(w) \int d\mu(A)f(A).
\end{equation}
One can prove positivity by conf\/irming $\omega(a^*a)= \int d\mu(A)|\sum_{i=1}^nf_i(A)|^2\ge0$. It will soon be important that this only requires that $d\mu$ is invariant under $\mathcal W$, but {\it{not}} that the action $\alpha_{.}$ of $\mathcal W$ on $C(\mathcal A)$ is by pull-backs.

To apply the def\/inition (\ref{equ:character-state}), we have to f\/ind a character of the noncommutative group of Weyl operators $\mathcal W$, which seems hopeless because $\mathcal W$ is noncommutative and inf\/inite dimensional. We thus consider a slightly dif\/ferent group $\mathcal W^\prime$. For this we f\/ix a maximal commutative subgroup~$\tau(x)$ of~$SU(2)$ at each point in~$x\in\mathbb X$ and for each surface~$S$ consider only the Weyl operators $w_{\Lambda_\tau,S,\sigma}=w_{\Lambda,S,\sigma}$ where $\Lambda(x)$ takes values in $\tau(x)$. This group is too small to construct \lqg. To be able to construct a version of \lqg, we consider the standard \lqg area operators~$\hat A(S)$ for each surface $S$. We now use denseness of the gauge-variant spin network functions~$T$ in~$C(\mathcal A)$ and consider the action of exponentiated area operators $w_{\lambda,S,A}=e^{i\lambda \hat A(S)}$, which act by multiplication of a phase\footnote{Notice that the action of the area operator on gauge-invariant spin-network functions is much more complicated, because one has to choose a recoupling basis. It follows that a gauge-invariant spin network containing vertices of valence four or more can not be a simultaneous eigenstate of all area operators.}  $e^{i\lambda \phi(S,T)}$ on the gauge-variant spin network functions. We now observe that the action of the area operators leaves $d\mu_{AIL}$ invariant. This is easily seem by expanding an element $f$ of $C(\mathcal A)$ in terms of spin network functions $f=f_o 1+(f-f_o 1)$ and noticing that $\int d\mu(A) f(A)=f_o$ and that the phase of the constant spin network function $1$ vanishes for all area operators. As in the case of the standard Weyl algebra, we have to include quasi-surfaces in the group $\mathcal W^\prime$ that is generated by the Weyl operators $w_{\lambda_\tau,S,\sigma}$ and $w_{\lambda,S,A}$. It turns out that all group elements of $\mathcal W^\prime$ leave $d\mu_{AIL}$ invariant and commute mutually. We denote the modif\/ied Weyl algebra by $\mathfrak A_o^\prime$.

To construct a character $\chi$ on $\mathcal W^\prime$, we use a densitized inverse triad $E$ describing a smooth classical geometry on $\mathbb X$ and def\/ine for 2-dimensional natural surfaces $S$
\begin{equation*}
   \chi_E(w_{\Lambda_\tau,S,\sigma_{nat.}}) = e^{i\int_S \Lambda_\tau.E}, \qquad
   \chi_E(w_{\lambda,S,A}) = e^{i\lambda A_E(S)},
\end{equation*}
where $A_E(S)$ denotes the classical area of $S$ calculated from $E$ and where
\[
(\Lambda_\tau.E)(x)=\Lambda(x)\tau_I(x)E^I(x)
 \]
 denotes a 2-form. All other elements of $\mathcal W^\prime$ are mapped to $1$. Using $\chi_E$ in the state def\/inition~(\ref{equ:character-state}) to def\/ine the state $\omega_E$, we see that we can construct a GNS representation of~$\mathfrak A_o$ for every classical geometry~$E$ on~$\mathbb X$.

Let us now construct the GNS-representations for $\omega_E$. To calculate the Gel'fand ideal, we observe that an element $a$ of $\mathfrak A_o$ lies in the Gel'fand ideal of $\omega_E$ if and only if $\kappa_E(a)$ is in the Gel'fand ideal of $\omega_o$, where the linear map $\kappa_E$ is def\/ined as $\kappa_E(f\circ w):=\chi_E(w) f \circ w$. The elements of the Gel'fand ideal of $\omega_o$ are of the form $a-f(a)$, where $f(a)\in C(\mathcal A)$ and applying $\kappa_E$ does not change this form. It follows that we can choose the image of $\eta_{\omega_E}$ takes values in $C(\mathcal A)$, so we can construct a ``Schr\"odinger'' representation analogous to the standard representation of \lqg. Moreover, the gauge-variant spin network functions form a dense orthogonal set in $\mathcal H^o_{\omega_E}$ and the trivial spin network function $1$ provides a cyclic vacuum vector $\Omega_E$. The expectation values for spatial geometric operators (i.e.\ the elements $w\in\mathcal W^\prime$) are given by the classical values calculated from the geometry $E$:
\begin{equation*}
 \langle \Omega_E,\pi_{\omega_E}(w)\Omega_E\rangle_{\omega_E}=\omega_E(w)=\chi_E(w).
\end{equation*}
We have thus constructed a representation of $\mathfrak A_o^\prime$ where the vacuum expectation values of all geometric operators coincide with classical expectation values for some spatial geometry~$E$. Moreover, it follows from direct calculation that $\Omega_E$ is an eigenstate of $\mathcal W^\prime$.

The construction of $\mathfrak A_o^\prime$ required us to restrict ourselves to $\mathcal W^\prime$ which does not allow to reconstruct all f\/lux operators of \lqg. It thus seems at f\/irst sight that one may not be able to construct all geometric operators from $\mathcal W^\prime$. However, since the classical area element contains enough information to reconstruct the metric (at each point one needs to consider six independent conormals), one can use these classical relations to construct quantum geometry operators purely from area operators. We thus see that the restriction to $\mathcal W^\prime$ does not restrict the number of available geometric operators of \lqg. Let us conclude this discussion with the remark that any discussion of the ``right'' quantization of geometric operators based on the quantization of classical expressions may be completely moot; ultimately geometry is determined by the dynamics of test matter living on it, e.g.\ by considering the correlation functions of a free f\/ield on a ``f\/ixed'' quantum geometry. The resulting notion of geometry at very short distances can in general not be expected to precisely coincide with the straightforward quantization of a classical relation between geometric operators.

The introduction of a geometric background may at f\/irst sight be seen as an obstruction to the implementation of bundle automorphisms (spatial dif\/feomorphisms and gauge transformations), because for any surface $S$ and any dif\/feomorphism $\phi$, if dif\/feomorphisms are implemented unitarily, we have
\begin{equation*}
 A_{E}(\phi(S))=\omega_E(U_\phi^* A(S)U_\phi)=\langle U_\phi \Omega_E,A(S) U_\phi \Omega_E\rangle_{\omega_E},
\end{equation*}
where $A_{E}(S)$ denotes the classical area of $S$ given geometry $E$. This can not be realized in~$\mathcal H_{\omega_E}$. The remedy for this problem is to read this as a def\/inition of $U_\phi \Omega_E$, which implies $U_\phi \Omega_{E}=\Omega_{\phi E}$. For this action to be unitary, we have to consider a Hilbert space~$\mathcal H^\phi_{\omega_E}$ that is the direct sum of all~$\mathcal H_{\omega_{E^\prime}}$, where~$E^\prime$ is in the dif\/feomorphism orbit of $E$. The unitary implementation of gauge transformations works analogously. One ends up with a huge direct sum Hilbert space~$\mathcal H_{G(E)}$, where~$G(E)$ denotes the orbit of~$E$ under bundle automorphisms. This is completely analogous to what is done in the case of the holonomy-f\/lux algebra, see \eqref{eq_dirsum} and below.  Applying the group averaging procedure to mod out the bundle morphisms gives a~gauge- and dif\/feomorphism-invariant Hilbert space~$\mathcal H_E$ that is spanned by gauge-invariant spin network functions that are embedded into a background geometry modulus isomorphisms of the background geometry~$G(E)$.

To conclude this section we remark that one can incorporate matter f\/ields in this construction as well. This was explicitly done for a scalar f\/ield that could play the role of an inf\/laton.

\subsection{Physical interpretation and comparison}

In this section, we would like to make some comments regarding the physical interpretation of the results we have sketched above, as well as the comparison between holonomy-f\/lux algebra and Weyl algebra approach.

\looseness=1
\emph{Essential geometry.} We saw that both the Lie algebra as well as the $C^*$-algebra construction lead to a representation of the algebra of \lqg where the vacuum vectors $\Omega_E$ were  eigenvectors of spatial geometry with eigengeometries prescribed by $E$. To interpret the vectors of $\mathcal H_{\omega_E}$, let us consider how one can reconstruct the ``background geometry'' $E$ from $v\in \mathcal H_{\omega_E}$. For this we def\/ine the excess of a state $v$ as the smallest countable set of zero and one-dimensional analytic sub-manifolds of the spatial hypersurface, such that the removal of any further zero- or one-dimensional analytic submanifold does not change any of the expectation values of two-dimensional geometric operators. Now def\/ine the {\it essential geometry} of a state~$v$ as the smooth geometry $E$ that is reconstructed from the expectation values of the two-dimensional geometric operators after the removal of the excess of~$v$. Notice that the excess of a state $v$ removes all \lqg excitations, since these are supported at most on a countable number of piecewise analytic edges; the resulting essential geometry is thus precisely the geometry~$E$ that we used to construct~$\omega_E$.

\emph{Background coupling.} Since operators that measure the essential geometry can be def\/ined solely from operators that are already in $\mathcal W^\prime$, it is natural to include these ``background''-operators. These operators $\tilde E_\tau(x)$, $|\tilde E|(x)$ act on a state $T\Omega_{E_o}$ with essential geometry $E_o$ as $(\tilde E_\tau(x),|\tilde E|(x))T\Omega_{E_o}=(E_{o,\tau}(x),|E_o|(x))T\Omega_{E_o}$. This is redundant if one considers a representation on one summand $\mathcal H_{\omega_E}$, since these operators are proportional to the unit operator and can thus be identif\/ied with multiplication by numbers. However, if one considers a representation on a direct sum $\mathcal H_{G(E)}$ then the background geometry operators are not proportional to the unit operator. Moreover, the transformation properties of these operators under bundle automorphisms is induced. Let us now consider a gauge-invariant operator $\hat O$ constructed from holonomy matrix elements and background geometry operators and consider the action on $\Omega_E$ as an element of $\mathcal H_{G(E)}$. Let us now apply the group averaging procedure for the bundle automorphisms to the resulting element $\hat O \Omega_E$. Due to the nontrivial transformation properties of the background geometry operators, one can not expand the resulting vector in gauge-invariant spin network functions, but one has to allow for gauge-invariant couplings of the spin network to the background geometry as well. We will hint towards a possible physical interpretation of this possible background coupling when we discuss the nondegenerate vacuum representations in the context of ef\/fective f\/ield theory.

\emph{Hamiltonian constraint quantization.} It is interesting to see whether the Hamiltonian constraint can be quantized in the new representations. On the one hand, one can try the standard techniques. In the $\hfalg$ approach, this can be carried out with a surprising result. In the standard approach \cite{Thiemann:1996aw}, a regularization of the Hamiltonian constraint is def\/ined on $\mathcal{H}_{\text{AL}}$ in terms of holonomies and the volume operator $V^{\vac}$. It can be checked that this construction can be repeated for each of the $\mathcal{H}_{\bg{E}}$ and hence on $\mathcal{H}_{G(\bg{E})}$. Thus the regularizations of the Hamilton constraint are well def\/ined, and solutions can be sought. The action of  $V$ is certainly modif\/ied in the new representations, but the volume operator only enters in terms of \emph{commutators} $[h,V]$ with holonomies or $[H_{\text{euc}},V]$ with the Euclidean Hamiltonian constraint.
These commutators are not modif\/ied. Therefore
\begin{equation*}
(f,\bg{E}| \widehat{H}(N) = 0 \Longleftrightarrow  (f,\bg{E'}| \widehat{H}(N) = 0
%\label{eq:}
\end{equation*}
for \looseness=1 $\bg{E}$, $\bg{E'}$ not necessarily gauge related. In particular, any solution over the standard vacuum is a solution over non-trivial backgrounds. The meaning of this observation is not clear.

On the other hand, with a non-degenerate background one has more possibilities available for the quantization of the Hamilton constraint. Essentially, the inverse volume problem is gone, and the quantization of the functional dependence on $E$ in the constraint gets trivialized. The problems with quantizing curvature, on the other hand, remain. How these observations can be used for a new quantization of the constraint is currently under study.

\emph{$\hfalg$ vs.\ $\mathfrak{A}$.} Although the holonomy-f\/lux representation theory together with correct domain assumptions and Fleischhack's Weyl algebra are closely related, there is a subtle dif\/ference. It arises, because the holonomy-f\/lux commutation relations do not close as a Lie algebra, because the commutator of two f\/lux operators (i.e.\ through a natural surface) is proportional to a quasi-f\/lux (i.e.\ through a quasi-surface). Regularity of the representation of the $C^*$-algebra ensures that one has a representation of the holonomy-f\/lux algebra on a dense domain. The holonomy-f\/lux algebra which closes to a particular Lie algebra involving quasi f\/lux operators. However, there are more Lie algebras that contain the physically important holonomy-f\/lux algebra. One can in particular relax the assumption that commutators of the physically dubious quasi f\/luxes are required to reproduce their classical Poisson brackets. One obtains a richer representation theory if one relaxes this assumption. This is the technical reason why the nondegenerate representations can only be def\/ined for the restricted algebra $\mathcal W^\prime$ and not all of~$\mathcal W$, while the analogous restriction is not necessary in the Lie-algebraic setting.

\section{Outlook}

The F/LOST-uniqueness result was a boost for the development of \lqg, because it basically taught us that there is a unique dif\/feomorphism-invariant kinematic ground state. Let us thus justify the nondegenerate representations by considering possible applications:

 1. {\it Effective Field Theory.} Ef\/fective f\/ield theory is a very valuable tool for the investigation of low-energy ef\/fects of a quantum f\/ield theory, that can be used even if the high energy theory is unknown. Representations with nondegenerate background geometry can be used to study ef\/fective f\/ield theory of \lqg f\/luctuations in a semiclassical background. The nondegenerate background geometry is particularly useful since its presence allows one to use e.g.\ the background metric to construct coarse graining operations for \lqg f\/luctuations. The idea is that loop structures that are smaller than a given scale w.r.t.\ the background metric are integrated out and are ef\/fectively described as shift in the background metric after a coarse graining step. This may open the door for the application of Wilsonian renormalization methods to \lqg. Such a procedure would interpolate between ``bumpy'' Loop Quantum Geometries in the UV and smooth background geometries in the IR. This avoids many of the problems associated with the geometric interpretation of spin network states.

 The asymptotic safety program by Reuter and collaborators does not have this problem, because it is formulated entirely in a continuum f\/ield theory setting. The idea behind this prog\-ram is to track the change of the ef\/fective action as one integrates out quantum f\/luctuations at a~momentum scale $k$. This program has accumulated nonperturbative evidence for the existence a~non-Gaussian f\/ixed point and it follows from simple dimensional analysis that the ef\/fective dimension at the non-Gaussian f\/ixed point is~2 while the ef\/fective dimension at low energies is~4. This suggested that the fundamental theory should have 1-dimensional excitations of space, which matches with spin network excitation in LQG. This has lead to speculations that LQG (with degenerate vacuum) may be able to provide the fundamental description of the f\/ixed point theory.

 A concrete implementation of this scenario is however elusive. A main problem is that a~coarse graining procedure that starts with a ``fundamental'' Ashtekar--Lewandowski state and ends up with a state with ef\/fective nondegenerate background geometry is missing. This is partly due to the aforementioned problem of f\/inding the continuum geometry that ``best approximates'' the loop geometry while smoothing out short distance bumpiness. It is thus at the moment not possible to depart from LQG (with degenerate vacuum) in the UV and end up with an ef\/fective continuum theory in the IR.

 Let us now return to background couplings in light of an ef\/fective f\/ield theory interpretation. Consider a spin network state whose graph is a regular lattice with small nonvanishing spin quantum numbers on all edges of the lattice plus a few nonlocal links with nonvanishing spin quantum numbers. Let us furthermore assume a coarse graining procedure that absorbs the entire lattice  as a shift of the background geometry, so only the nonlocal links are left after coarse graining. It has been suggested that the endpoints of nonlocal links should appear like matter to a local observer. If this assertion holds then one should interpret the background couplings as Loop Quantum Geons.

 2. {\it Reinterpretation of Loop Quantum Cosmology.} Loop Quantum Cosmology is generally understood as a toy model that incorporates important features of \lqg, but a precise mathematical relation is not established. Nondegenerate representations allow for a dif\/ferent interpretation where standard Loop Quantum Cosmology is interpreted as the dynamics of a homogeneous background geometry. This could be used to couple arbitrary Loop Quantum f\/luctuations to standard Loop Quantum Cosmology.

3. {\it Further Generalizations.} The main technical problem in constructing the above states is to prove positivity of the GNS functional. The positivity proof for the GNS functionals allows for several generalizations, which are not explored. It may be particularly useful to investigate states that are invariant under certain f\/lows. This may avoid ``quantization'' by providing a completely mathematical selection criterion for Loop Quantum dynamics as unitarily implementable f\/lows.

Let us conclude this review with a provocative thought: ``The nondegenerate representations may hint what the essence of quantum gravity is.'' Let us explain what we mean: Many of the nice features of \lqg are due to the fact that the kinematic Hilbert space can be constructed as the closure of a set of normalized and mutually orthogonal eigenstates of spatial quantum\footnote{We include the attribute quantum, because of the noncommutativity of Loop Quantum Geometry operators on the gauge-invariant Hilbert space.} geometry. E.g.\ the unitarity of the natural action of spatial dif\/feomorphisms as well as the ``discreteness'' of spatial geometric operators is a direct consequence. This is particularly transparent in the nondegenerate representations. It is thus not unreasonable to attempt new approaches to quantum gravity that start with a Hilbert space that is the completion of mutually orthogonal normalized eigenstates of spatial geometry and implement exponentiated momentum operators as pull-backs under translations in the geometrodynamic conf\/iguration space.

\subsection*{Acknowledgements}

HS gratefully acknowledges partial support through the Spanish MICINN Project No.\ FIS2008-06078-C03-03. Research at the Perimeter Institute is supported in part by the Government of Canada through NSERC and by the Province of Ontario through MEDT.

\pdfbookmark[1]{References}{ref}
\LastPageEnding

\end{document}